\documentclass[12pt,preprint, epsfig]{aastex}

\shorttitle{Local Group distances and publication bias. I. The LMC}
\shortauthors{Richard de Grijs et al.}

\begin{document}

\title{Clustering of Local Group distances: publication bias or
  correlated measurements? I. The Large Magellanic Cloud}

\author{
Richard de Grijs\altaffilmark{1,2},
James E. Wicker\altaffilmark{3}, and 
Giuseppe Bono\altaffilmark{4,5}
}

\altaffiltext{1} {Kavli Institute for Astronomy and Astrophysics,
  Peking University, Yi He Yuan Lu 5, Hai Dian District, Beijing
  100871, China} 
\altaffiltext{2} {Department of Astronomy, Peking University, Yi He
  Yuan Lu 5, Hai Dian District, Beijing 100871, China}
\altaffiltext{3} {National Astronomical Observatories, Chinese Academy
  of Sciences, 20A Datun Road, Chaoyang District, Beijing 100012,
  China} 
\altaffiltext{4} {Dipartimento di Fisica, Universit\`a di Roma Tor
  Vergata, via Della Ricerca Scientifica 1, 00133, Roma, Italy}
\altaffiltext{5} {INAF, Rome Astronomical Observatory, via Frascati
  33, 00040, Monte Porzio Catone, Italy}

\begin{abstract}
The distance to the Large Magellanic Cloud (LMC) represents a key
local rung of the extragalactic distance ladder. Yet, the galaxy's
distance modulus has long been an issue of contention, in particular
in view of claims that most newly determined distance moduli cluster
tightly---and with a small spread---around the ``canonical'' distance
modulus, $(m-M)_0 = 18.50$ mag. We compiled 233 separate LMC distance
determinations published between 1990 and 2013. Our analysis of the
individual distance moduli, as well as of their two-year means and
standard deviations resulting from this largest data set of LMC
distance moduli available to date, focuses specifically on Cepheid and
RR Lyrae variable-star tracer populations, as well as on distance
estimates based on features in the observational Hertzsprung--Russell
diagram. We conclude that strong publication bias is unlikely to have
been the main driver of the majority of published LMC distance
moduli. However, for a given distance tracer, the body of publications
leading to the tightly clustered distances is based on highly
non-independent tracer samples and analysis methods, hence leading to
significant correlations among the LMC distances reported in
subsequent articles. Based on a careful, weighted combination, in a
statistical sense, of the main stellar population tracers, we
recommend that a slightly adjusted canonical distance modulus of
$(m-M)_0 = 18.49 \pm 0.09$ mag be used for all practical purposes that
require a general distance scale without the need for accuracies of
better than a few percent.
\end{abstract}


\section{Introduction}
\label{intro.sec}

The distance to the Large Magellanic Cloud (LMC) is a key stepping
stone in establishing an accurate extragalactic distance ladder. The
LMC is the nearest extragalactic environment that hosts statistically
significant samples of the tracer populations that are commonly used
for distance determinations, including Cepheid and RR Lyrae variable
stars, eclipsing binaries (EBs), and red-giant-branch (RGB) stars, as
well as supernova (SN) 1987A, among others. These could thus
potentially link the fairly well-understood local (solar-neighborhood
and Galactic) tracers to their counterparts in more distant and more
poorly resolved galaxies. In fact, at a distance of approximately 50
kpc, the LMC represents the only well-studied environment linking
Galactic distance tracers to those in other large spiral and
elliptical galaxies at greater distances, including M31 at a distance
of $\sim 750$--780 kpc or a distance modulus of $(m-M)_0 =
24.38$--24.47 mag (e.g., Freedman et al. 2001; McConnachie et
al. 2005). Yet, despite the plethora of studies claiming to have
determined independent distance measurements to the LMC, lingering
systematic uncertainties remain. This has prompted significant concern
in the context of using the LMC distance as a calibrator to reduce the
uncertainties in the Hubble constant (cf. Freedman et al. 2001;
Schaefer 2008; Pietrzy\'nski et al. 2013). It has also led to
persistent claims of ``publication bias'' affecting published
distances to the galaxy (cf. Schaefer 2008, 2013; Rubele et al. 2012;
Walker 2012).

In general, publication bias is the tendency of researchers to publish
results that conform to some extent to the norm, while ignoring
outputs that may be of low(er) significance or that deviate
significantly from what is considered common knowledge in the relevant
field. In other words, the strongest or most significant results are
included for publication, while the rest of a presumably much larger
set of results remain unseen. This also means that this effect is
notoriously difficult to correct for, because the underlying null
results are usually not published. The phenomenon of publication bias
is well-known to occur in statistics and among medical trials (e.g.,
Sterling 1959; Rosenthal 1979; Begg \& Berlin 1988; Naylor 1997; Stern
\& Simes 1997; Sterne et al. 2000), where it could have potentially
devastating effects on people's lives, or lead to ineffectual or even
counterproductive treatments.

Liddle (2004) explains that ``[p]ublication bias comes in several
forms, for example if a single paper analyses several parameters, but
then focusses attention on the most discrepant, that in itself is a
form of bias. The more subtle form is where many different researchers
examine different parameters for a possible effect, but only those
who, by chance, found a significant effect for their parameter,
decided to publicize it strongly.'' Publication bias has also been
claimed to occur in various fields related to astrophysics and
cosmology, where in some cases efforts have also been made to correct
for these effects (see, e.g., Slosar \& Seljak 2004; Slosar et
al. 2004; Schaefer 2008, 2013; Vaughan \& Uttley 2008; Bailer-Jones
2009; Sternberg et al. 2011; Foley et al. 2012).

In the context of the present paper, Schaefer (2008) focused his
analysis on published LMC distance determinations. He specifically
addressed the question as to whether or not the publication of the
final results of the {\sl Hubble Space Telescope} ({\sl HST}) Key
Project on the Extragalactic Distance Scale (Freedman et al. 2001) had
resulted in an unwarranted tightening up of the LMC's distance
scale. He considered as possible causes of such a tightening
correlations among published results, widespread overestimation of
uncertainty ranges, bandwagon effects, or a combination of these
scenarios. He concluded with a warning that the community would do
well to be vigilant and redress the effects of publication bias, which
he considered the most likely cause of the clustering of LMC distance
measurements he reported to have occurred during the period from 2002
until June 2007.

Upon careful examination, however, we realized that Schaefer's (2008)
analysis---as well as his subsequent persistence in support of his
2008 conclusion that publication bias may have severaly affected the
body of LMC distance measurements (e.g., Schaefer 2013)---was based on
a number of simplifying assumptions:
\begin{enumerate}
\item He concludes that the uncertainties in the post-2002 distance
  moduli are not distributed according to a Gaussian distribution,
  which he flags as a problem. However, in such a scenario, the
  underlying assumptions are that (i) the pre-2001 values were, in
  fact, distributed in a Gaussian fashion (they are not, however, as
  we will show in Section \ref{lmcdist.sec}) and (ii) conditions were
  comparable before and after the benchmark date. This latter
  assumption is likely also too simplistic, as we will argue in the
  context of Cepheid-based distance determinations in Section
  \ref{cepheids.sec}. We recommend---and pursue in this paper---a more
  detailed analysis of the individual distance moduli contributing to
  the overall trends observed to assess whether or not publication
  bias truly is to blame.
\item Schaefer (2008) based his results on published values and their
  uncertainties; the latter are, however, predominantly statistical
  uncertainties and the majority do {\it not} include systematic
  errors. Only a few authors include the systematic errors affecting
  their LMC distance estimates, however. In Section
  \ref{statistics.sec} we apply statistical tools to assess whether
  the distance moduli based on different tracer populations are
  statistically consistent with the ``canonical'' distance modulus and
  the recently published geometric distance based on late-type EB
  systems (Pietrzy\'nski et al. 2013). We also compare the consistency
  among a number of different tracers and the entire body of distance
  measurements (see Sections \ref{bias.sec} and
  \ref{conclusions.sec}).
\item The conclusions reached by Schaefer (2008) are, in essence,
  based on application of a statistical Kolmogorov--Smirnov (KS) test,
  assuming a Gaussian distribution of LMC distance measurements, to a
  data set that should not {\it a priori} be expected to be
  distributed in a Gaussian fashion. Astrostatisticians have become
  more vocal in recent years in their opposition to the use of KS
  tests in astronomy if not done with due caution (e.g., Feigelson \&
  Babu 2013). KS tests are only applicable to samples that consist of
  independent and identically distributed values. In the context of
  LMC distance measurements, both conditions are violated. In this
  paper we will show that the close match between subsequent LMC
  distance determinations is most likely owing to the use of highly
  non-independent tracer samples, analysis methods, and calibration
  relations.
\item As Schaefer (2008) points out himself, his database of LMC
  distance measurements is incomplete. He declares that this does not
  affect his inferences, but we found that gaps in the data set may,
  in fact, hide the presence of correlations among subsequent
  publications (cf. Section \ref{bias.sec}). For the analysis
  presented in this paper, we have collected the most complete and
  comprehensive database of published LMC distance moduli to
  date,\footnote{Schaefer (2008) lists 44 articles containing as many
    new LMC distance moduli published between July 2001 and April
    2007. In that same period, our database includes 49 articles with
    a total of 67 new LMC distance determinations. Note that for this
    comparison we did not consider the final entry in Schaefer's
    (2008) list, which at the time of his publication had just
    appeared on the arXiv preprint server
    (http://www.arXiv.org/archive/astro-ph), but which did not appear
    in the printed literature until June 2008 (Ngeow \& Kanbur 2008).}
  so that our results will not be unduly affected by ``gaps'' in the
  coverage of our metadata.
\end{enumerate}

These concerns, combined with the significantly longer period
(compared with that accessible by Schaefer 2008) that has elapsed
since Freedman et al.'s (2001) seminal paper, prompted us to embark on
a detailed (re-)analysis of the full set of LMC distance
determinations, claimed by many of their authors to be based on
independent approaches (but see Section \ref{bias.sec}). The primary
goal of the analysis presented in this paper is to explore the reasons
behind the apparent tightening of the biennially (two-year) averaged
distance moduli and the associated decrease in their standard
deviations during specific periods of time. We aim at exploring
whether ``publication bias'' is likely to play a significant role in
driving this behavior or whether other effects may be at work. The
longer time span we have access to compared with previous work also
allows us to verify whether any clustering of data points persisted
beyond the period range of Schaefer's (2008) analysis and whether new
clusters of data points may have materialized. Our detailed analysis
of the complete body of published LMC distance moduli from 1990 until
the end of 2013, both in full and as a function of distance tracer, is
ideally suited to derive statistically robust constraints on the most
appropriate mean distance modulus (projected to the LMC's center) and
its uncertainties for future use ({\it modulo} the quality of the
individual determinations).

This paper is organized as follows. In Section \ref{data.sec}, we
present the full compilation of LMC distance moduli published between
1990 and the present time. Section \ref{lmcdist.sec} addresses general
trends in the LMC distance moduli with time, while in Section
\ref{bias.sec} we consider such trends for individual distance tracers
and discuss the independence (or lack thereof) of the results. We
discuss the statistical basis of our analysis in Section
\ref{statistics.sec}. In Section \ref{conclusions.sec} we place these
results in a more general context, and we conclude with our
recommendations of the most suitable distance modulus for common use,
which naturally results from the analysis presented here. In Paper II
(de Grijs et al. 2014) we apply a similar analysis to our compilation
of the equivalent sets of distance measurements for M31, M33, and a
number of dwarf galaxies associated with the M31 system (and slightly
beyond).

\section{LMC Distance Measurements, 1990--2013}
\label{data.sec}

We compiled an extensive database of published distance measurements,
following but significantly expanding upon Schaefer's (2008)
database. We used the compilations of Gibson (2000), Benedict et
al. (2002a), Clementini et al. (2003), Steer \& Madore (2007),
Schaefer (2008), and de Grijs (2011; his table 1.1) as our
basis. Schaefer (2008) stated specifically that his final compilation
was not necessarily complete. Therefore, we carefully checked which of
the nearly 16,000 papers published between 1 January 1990 and 31
December 2013 and associated in the NASA/Astrophysics Data System
(ADS) with the object keyword ``LMC'' presented new distance
determinations to the galaxy.

Our final compilation contains 233 separate distance determinations
published from March 1990 until and inclusive of December 2013. In
addition to recording the month and year of publication, we compiled
the extinction-corrected distance moduli, their statistical
uncertainties and---where available---also the systematic errors. Only
47 authors published their systematic uncertainties separately, with
an additional four papers specifying that their published error bars
include the systematic uncertainties. For the remaining 182 LMC
distance measurements, the uncertainties refer to the statistical
errors only. Our overall compilation, in order of publication date as
well as sorted by stellar population tracer, is available from
http://astro-expat.info/Data/pubbias.html (all sources that were
available at the time of our last update are listed, including those
published in early 2014; our analysis is restricted to distance moduli
published by 31 December 2013, however). The data tables include full
bibliographic information and direct links to the source materials.

The database is predominantly based on distance measurements taken
from peer-reviewed articles, although we include a total of 19
distance moduli that were published in conference proceedings. Seven
of these latter values were based on Cepheids (from four different
publications), six on RR Lyrae variables (from three conference
papers) and three on geometric distance determinations (two using SN
1987A and one based on an early-type EB system). The remaining
distance moduli published in conference proceedings were based on
observations of M supergiants, Mira variable stars, and the planetary
nebulae luminosity function (one determination for each method). We
opted to include these values, because individual researchers will
likely have checked the recent literature---including conference
proceedings---for confirmation of their newly derived values. In
addition, only two conference articles were followed up by
publications in the peer-reviewed literature in the period from 1990
until the present time (Popowski \& Gould 1998 vs 1999; Dall'Ora et
al. 2004a vs 2004b). The distance moduli published in both cases were
different between the earlier conference publication and the follow-up
peer-reviewed article. The Dall'Ora et al. (2004a) conference paper
was presented in May 2003, followed by a peer-reviewed article more
than a year later in July 2004. Popowski \& Gould (1998) used the Bono
et al. (2001) calibration, while their 1999 paper was based on the
updated Bono et al. (2003) calibration relations. We considered this
sufficiently distinct to warrant inclusion of both articles in our
database; the differences between the resulting distance moduli act as
a reminder of the systematic uncertainties involved.

Rather than combining individual values based on different assumptions
or input parameters (cf. Schaefer 2008), we opted to include all
(final) LMC distance measurements published in a given paper. Schaefer
(2008) argues (and we agree) that the range spanned by such
alternative values provides a valuable estimate of the systematic
uncertainty inherent to the distance determined, and that these values
are highly non-independent. It is nevertheless instructive to compare
the different values derived from varying one's assumptions based on
realistic boundaries. We will take into account any correlations among
results from a given study, as well as correlations between studies
based on similar (or the same) assumptions, when we discuss trends as
a function of year of publication for the different tracers
individually in Section \ref{bias.sec}.

Examples of correlated results based on small variations of the
underlying assumptions {\it in a given study} include
\begin{enumerate}
\item variations in the extinction corrections,\footnote{Reddening
  corrections have been and continue to be a major source of
  uncertainty: a number of different reddening maps are in common use
  (for a discussion, see Haschke et al. 2011), which has in part led
  to the long--short distance dichotomy discussed in Section
  \ref{lmcdist.sec}. This is why, specifically in the context of
  variable-star analyses, the focus has gradually shifted from the use
  of wavelength- and reddening-dependent period--luminosity relations
  to reddening-free Wesenheit relations (cf. Inno et al. 2013; Ripepi
  et al. 2013).} metallic\-ities/$\alpha$ abundances or $p$
  (``projection'') factors\footnote{Projection ($p$) factors are
  commonly used to convert radial to pulsation velocities.} assumed
  (e.g., Caputo 1997; Benedict et al. 2002a,b; Storm et al. 2006;
  Haschke et al. 2012);
\item use of a given type of calibration relation for a specific
  tracer but {\it for different wavelengths} (e.g., Di Benedetto 1994;
  van Leeuwen et al. 1997; Madore \& Freedman 1998; Groenewegen \&
  Oudmaijer 2000; Bono et al. 2002b; Dall'Ora et al. 2004a; An et
  al. 2007; Laney et al. 2012);
\item application of different calibration methods for a given tracer
  (e.g., Feast 1995; Luri et al. 1998; Bono et al. 2002b; Clementini
  et al. 2003; McNamara et al. 2007); and
\item differences in the exact shape of the ring associated with SN
  1987A (Gould \& Uza 1998).
\end{enumerate}

However, we also realized that, in many articles that report multiple
distance measurements, not only the assumptions underlying these
measurements differ, but also the tracers used to derive them or their
locations across the LMC's main body.\footnote{These articles include
  Reid (1998), Udalski (1998), Luri et al. (1998), Popowski \& Gould
  (1998, 1999), Feast (1999), Walker (1999), Carretta et al. (2000),
  Romaniello et al. (2000), Sakai et al. (2000), McNamara (2001),
  Benedict et al. (2002a)---and Benedict et al. (2002a vs
  2002b)---Clementini et al. (2003), Groenewegen \& Salaris (2003),
  Salaris et al. (2003), Rastorguev et al. (2005), Ngeow \& Kanbur
  (2008), Borissova et al. (2009), Subramanian \& Subramaniam (2010),
  Haschke et al. (2012), and Inno et al. (2013)} The combination of
multiple values offers one good insights into the systematic
uncertainties. However, we argue that inclusion of the individual
values is justified in this context, in particular because many of the
distance determinations published in subsequent papers are {\it also}
highly non-independent (see Section \ref{bias.sec}).

Despite these differences in the choices made, we concur with Schaefer
(2008) regarding the multitude of ``minor dilemmas'' that occur when
deciding which values to include in one's master database. In essence,
we followed his choices (as well as equivalent choices for distance
moduli not included in his database). Only in one case in common do we
differ in our approach. Schaefer (2008) comments on the
unrealistically small uncertainty on the LMC's distance modulus
published by Keller \& Wood (2006), despite the fact that these latter
authors provide a detailed discussion of the uncertainties affecting
their results. Upon close examination of the Keller \& Wood (2006)
result, we realized that they give the uncertainty on the mean of
their distribution of distance moduli, assuming a Gaussian
distribution (which is not really warranted given the actual
distribution of data values), instead of the distribution's spread
($\sigma$). Therefore, we have included the spread, which we
determined at $\sigma = 0.06$ mag for both of their published distance
moduli (Keller \& Wood 2002, 2006).

\section{Trends in Distance Determinations to the LMC}
\label{lmcdist.sec}

\begin{figure}
\begin{center}
\includegraphics[width=\columnwidth]{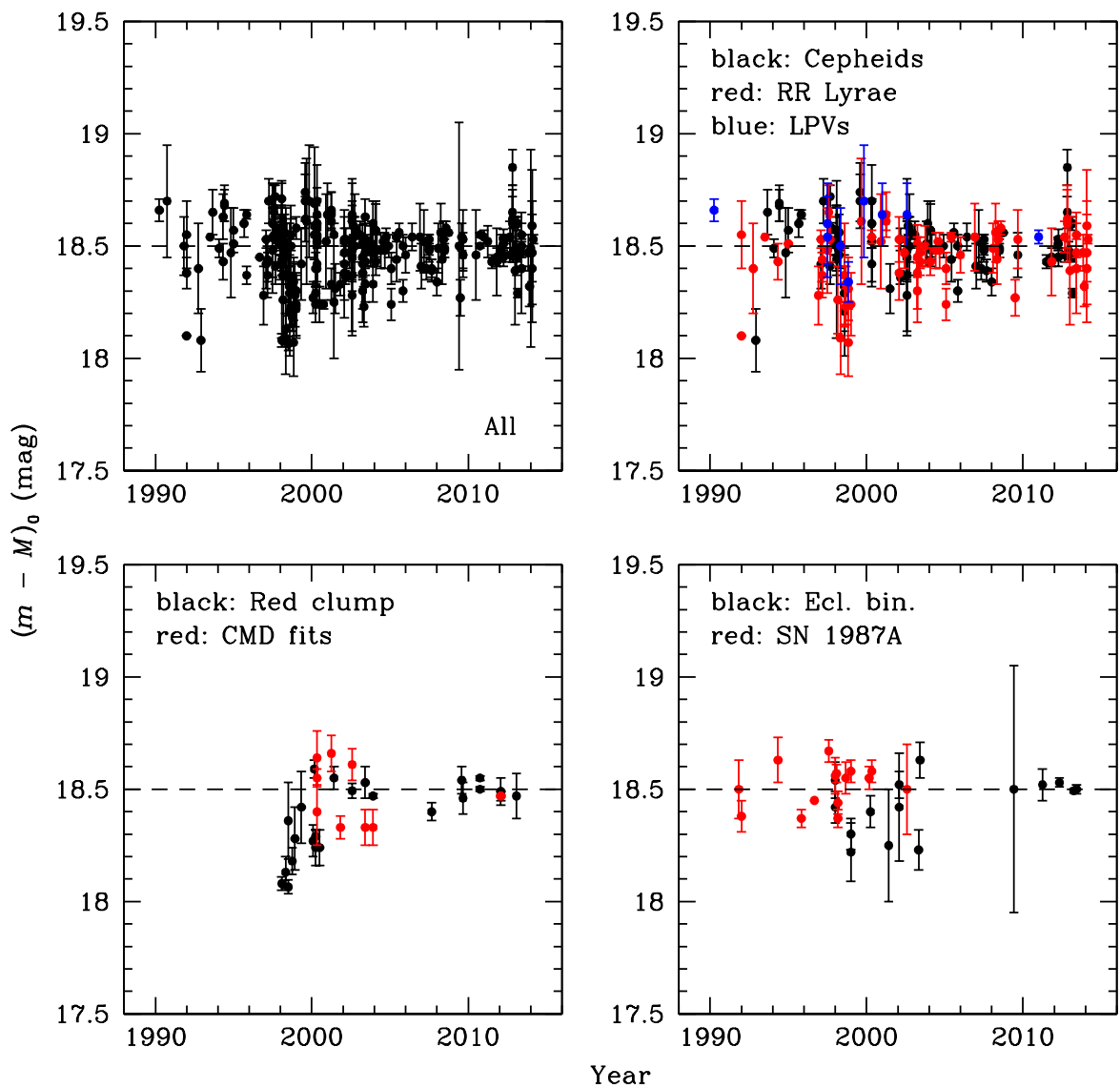}
\caption{Published extinction-corrected LMC distance moduli since 1990
  as a function of publication date (month), where possible centered
  on the galaxy's center. The horizontal dashed lines indicate the
  ``canonical'' distance modulus, $(m-M)_0 = 18.50$ mag (Freedman et
  al. 2001). LPVs: long-period variables; CMD: color--magnitude
  diagram; Ecl. bin.: eclipsing binaries; SN: supernova.}
\label{lmcdist.fig}
\end{center}
\end{figure}

Figure \ref{lmcdist.fig} displays the individual LMC distance moduli
published between 1990 and 2013. The top left-hand panel shows all
determinations, irrespective of distance tracer. The other three
panels provide more details as regards the behavior of specific
tracers with time. In all cases, except for the values determined by
Keller \& Wood (2002, 2006; see Section \ref{data.sec}), the error
bars included are those provided in the original papers. Where authors
distinguish between statistical and systematic uncertainties, we only
show the statistical uncertainties. (Note that, where available, the
statistical uncertainties reported in the individual articles are
specifically included in our online database.)

We made this choice for a number of reasons. First, the uncertainties
quoted for the majority of published LMC distance moduli are
statistical uncertainties. Showing only the statistical uncertainties
where we also have the systematic errors thus represents a cleaner
comparison among the distance values. Second, where multiple distance
moduli were determined in a given paper, the spread among the derived
values provides a good handle on the systematic uncertainties. Third,
in many cases, the effects of systematic uncertainties are unclear or
may have been underestimated (cf. Schaefer 2008). Systematic
uncertainties in the LMC distance modulus can come in a variety of
guises. They could be related to (i) one's zero-point calibration,
(ii) the functional form of the calibration relations (see Section
\ref{bias.sec}), (iii) Lutz--Kelker-type biases in parallax
measurements (see de Grijs 2011; his chapter 6.1.2), (iv) systematics
in the metallicity scale or extinction corrections adopted, or (v)
transformations between filter systems. They may also be due to
assumptions made to derive the underlying physical behavior or
geometry of one's tracer population (e.g., as for the SN 1987A ring
and the precise locus of its line emission; cf. Gould \& Uza 1998). An
important type of systematic uncertainty is immediately apparent from
inspection of the top left-hand panel of Figure \ref{lmcdist.fig},
which takes the form of the well-known ``long--short dichotomy.''

Particularly at early times during the period of interest on which we
focus here, authors would derive LMC distance moduli that were either
``short'' or ``long.'' Values straddled\newline $(m-M)_0 \sim 18.3$
mag and $\sim 18.6$ mag, respectively, hence leading to a dichotomy in
published LMC distance moduli prior to about 2002 (see the top
left-hand panel of Figure \ref{lmcdist.fig}). This is a typical
example of publication bias in astronomy (for discussions of the
long--short dichotomy, see Carretta et al. 2000; Sandage \& Tammann
2006). In this context, it is puzzling that Schaefer (2008) uses his
conclusion that the errors in the LMC distance moduli published in
2002--2007 (when the dichotomy had all but disappeared) deviate from a
Gaussian distribution as evidence that there must be a problem with
the overall {\it post-2002} data set. This implicitly assumes that the
post-2002 values are independent and distributed in a Gaussian
fashion; we will show that both assumptions are not justified once one
analyzes the data set in detail. Thus, while the 1997--2001 distance
uncertainties may indeed have been larger than expected from a
Gaussian distribution, this should not have been a surprise given that
the pre-2001 distance values exhibited a clear long--short dichotomy
owing to publication bias.

This dichotomy disappeared around 2002--2004 (but compare Dambis et
al. 2013 with the canonical LMC distance modulus for more recent
evidence of lingering systematics); yet, Schaefer (2008) interprets
the post-2002 behavior as a result of publication bias and he
perpetuates this view until the present time (Schaefer 2013). Here we
question that latter conclusion. In Section \ref{bias.sec}, we will
explore the trends for individual distance tracers so as to ascertain
the reasons for a number of statistically significant reductions in
the quoted error bars.

\section{Publication Bias or Correlated Methods?}
\label{bias.sec}

\begin{figure}
\begin{center}
\includegraphics[width=\columnwidth]{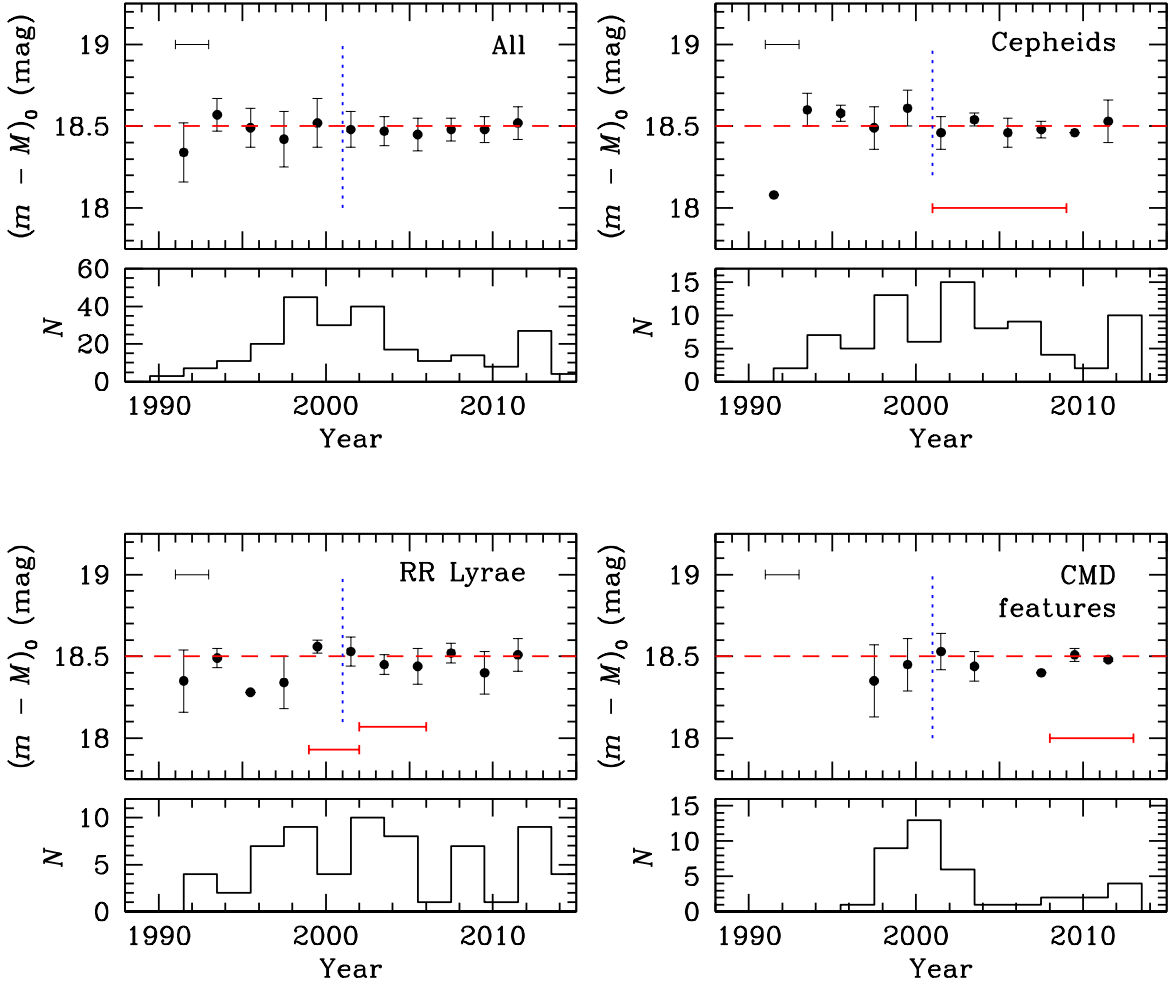}
\vspace{-1cm}
\caption{As Figure \ref{lmcdist.fig}, but averaged over two-year
  periods and divided into common distance tracers. The bottom panels
  associated with each main panel show histograms of the number of
  data values contributing to the two-year averages. The error bars
  represent the spread in (standard deviations of) the distance
  moduli; data points without error bars represent single entries
  during the relevant two-year period. The vertical blue dotted lines
  indicate the publication date of Freedman et al.'s (2001) canonical
  distance modulus. The horizontal error bars in the top left-hand
  corners of each main panel show the time period over which the data
  values have been averaged.}
\label{lmcdist2yr.fig}
\end{center}
\end{figure}

\begin{figure}
\begin{center}
\includegraphics[width=\columnwidth]{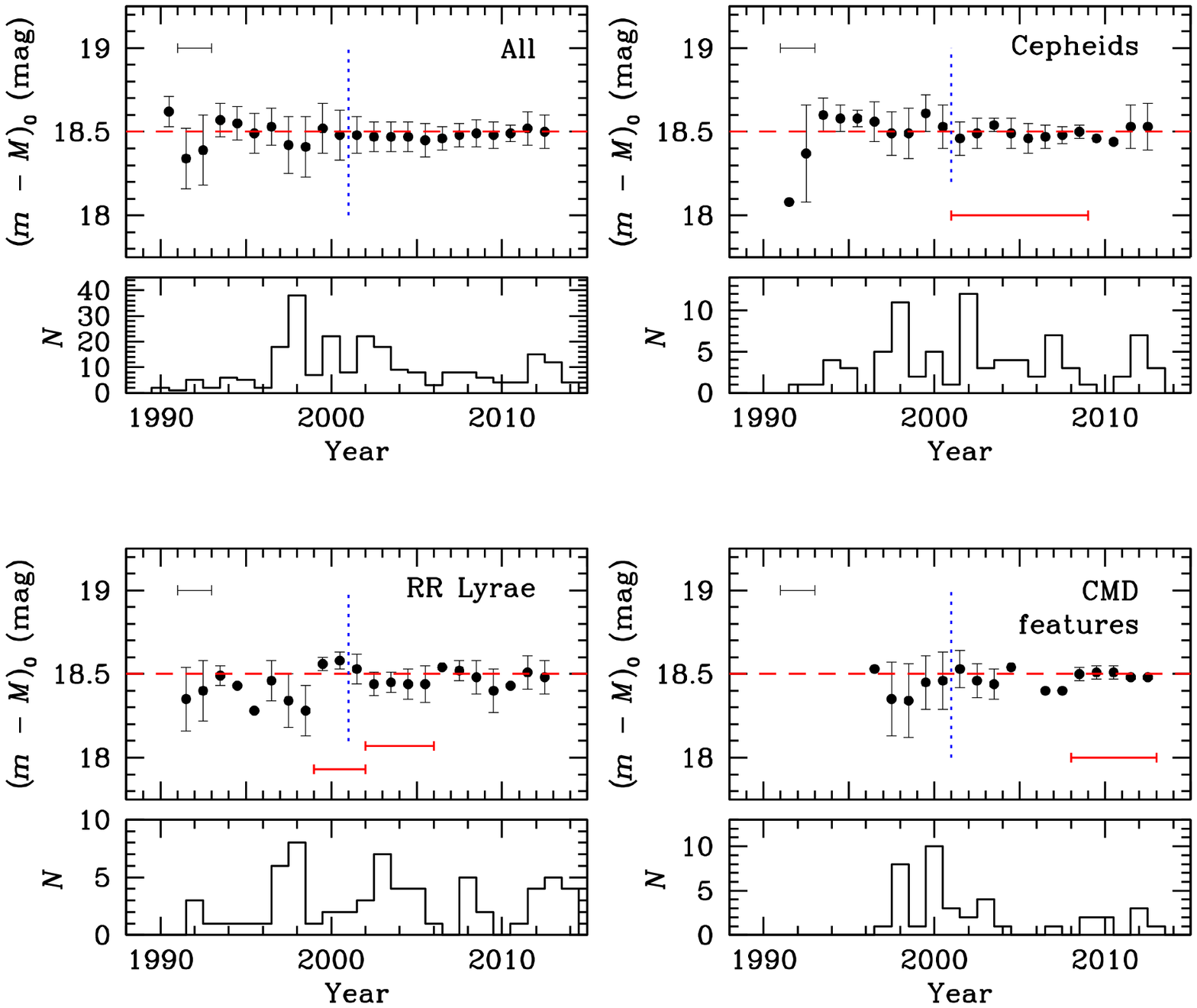}
\vspace{-1cm}
\caption{As Figure \ref{lmcdist2yr.fig} but for two-year {\it running}
  averages to highlight periods during which the spread in (i.e., the
  error bars associated with) the distance moduli exhibited sustained,
  statistically significant reductions. The red horizontal bars
  indicate the periods selected for further scrutiny, based on
  statistical considerations. The histograms show the year-on-year
  variations in numbers of published distance moduli.}
\label{lmcdist2run.fig}
\end{center}
\end{figure}

\begin{figure}
\begin{center}
\includegraphics[width=\columnwidth]{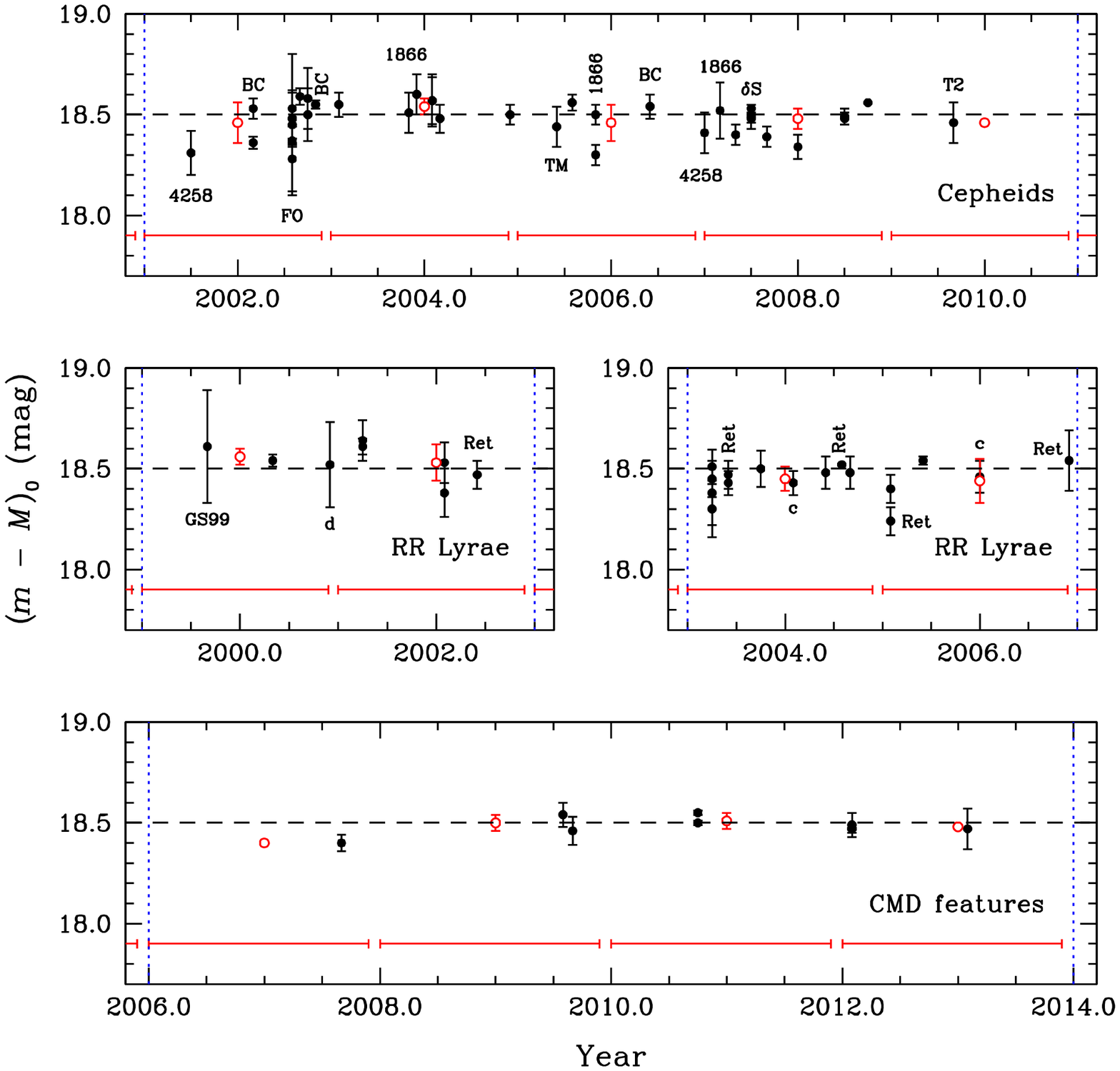}
\caption{Individual LMC distance moduli for zoomed-in date ranges
  identified on statistical grounds as characterized by small scatter
  and tight clustering of distance values. The red open circles
  represent the average values for the date ranges indicated by the
  horizontal red bars at the bottom of all panels. (top) Most
  Cepheid-based distance determinations are based on fundamental-mode
  pulsators from the full {\sc ogle ii} database, except where
  indicated. ``1866:'' cluster Cepheids in NGC 1886; ``4258:'' based
  on relative distance moduli between the LMC and NGC 4258; BC: bump
  Cepheids; $\delta$S: $\delta$ Scuti stars; FO: first-overtone
  pulsators; TM: triple-mode pulsators; T2: Type II Cepheids. (middle)
  ``c,'' ``d:'' distance moduli based on RRc (first-overtone) and RRd
  (double-mode) variables, respectively; GS99: Groenewegen \& Salaris
  (1999); Ret: distance moduli based on RR Lyrae stars in the
  Reticulum globular cluster. (bottom) All CMD-based distances are
  based on calibration of RC magnitudes.}
\label{indivdistLMC.fig}
\end{center}
\end{figure}

In Figures \ref{lmcdist2yr.fig}--\ref{stddev.fig} we show our metadata
in a number of different ways to quantify and explore the trends that
may be apparent in the actual distance moduli and their quoted
uncertainties.

Figure \ref{lmcdist2yr.fig} displays two-year averages of the
published LMC distance moduli for all tracers, as well as for those
tracers for which we have statistically significant numbers of
measurements. In particular, we show the results for the Cepheid and
RR Lyrae variable stars (80 and 60 entries, respectively), and for
distance tracers based on certain features in the color--magnitude
diagrams (CMDs) of LMC star clusters, for which we have collected 38
entries. Among the latter, we include results based on main-sequence
(MS)/MS turnoff (MSTO) fitting (Carretta et al. 2000; McNamara 2001;
Walker et al. 2001; Kerber et al. 2002; Groenewegen \& Salaris 2003;
Salaris et al. 2003), subdwarf fitting (Carretta et al. 2000), as well
as on fits to white dwarf cooling sequences (Carretta et al. 2000),
horizontal-branch (HB) stars (Gratton 1998; Reid 1998), the tip of the
RGB (TRGB; Salaris \& Cassisi 1997; Romaniello et al. 2000; Sakai et
al. 2000; Bellazzini et al. 2004), and the red clump (RC; 24
entries). The full bibliography, as well as the set of individual
values and their uncertainties for all distance tracers, are available
at http://astro-expat.info/Data/pubbias.html. The only tracers that we
do not examine individually in Section \ref{bias.sec} are the
long-period variable stars (9 entries), EBs (15 entries), SN 1987A (15
entries, published mostly between 1994 and 2000), novae (1 entry), the
planetary nebulae luminosity function (3 entries), and average values
based on multiple calibrations (7 entries). We will, however, return
to the geometric distance tracers, specifically to the EBs, in Section
\ref{conclusions.sec}.

For comparison, in Figures \ref{lmcdist2yr.fig} and
\ref{lmcdist2run.fig} we also show the level of the ``canonical'' LMC
distance modulus, $(m-M)_0 = 18.50$ mag (Freedman et al. 2001;
horizontal dashed lines), as well as its publication date (vertical
blue dotted lines). The error bars on the biennial average distance
moduli represent the standard deviations ($1 \sigma$ spreads) in the
data values. Aiming at highlighting periods of (statistically)
lower-than-expected uncertainties and/or smaller-than-expected scatter
among subsequent data points, we show two-year {\it running} averages
of the published LMC distance moduli in Figure
\ref{lmcdist2run.fig}. Inspection of the panels for the individual
distance tracers reveals a number of periods during which the average
values are {\it statistically} more tightly clustered and/or
associated with very small error bars (spreads). We will examine the
reasons underlying this behavior in the following subsections. In
Figure \ref{lmcdist2run.fig}, the red horizontal bars indicate these
periods of interest.

Specifically, for the Cepheid-based distance determinations, we
identified the period from 2001 to 2009 for further scrutiny; for the
RR Lyrae stars, we will consider the ranges 1999--2003 and 2003--2007,
and for the CMD features we will carefully examine the publications
leading to the distance moduli published in the period since 2006
until the present time. The individual data points in these date
ranges are shown in Figure \ref{indivdistLMC.fig}. We also examined
any trends in the (statistical) spreads as a function of publication
date: see Figure \ref{stddev.fig} and Section
\ref{conclusions.sec}. To determine whether the tight clustering of
data points is an artifact caused by publication bias or simply owing
to correlations among subsequent publications, we will examine both
the nature of the tracer samples and the calibration relations used.

\subsection{Cepheids}
\label{cepheids.sec}

The top right-hand panels of Figures \ref{lmcdist2yr.fig} and
\ref{lmcdist2run.fig} show a statistically significant reduction in
the uncertainties associated with the two-year averages between 2001
and 2009. During the entire period, the number of data points
contributing to the average values was at least five, so that the
average values and spreads are well-defined. This sudden reduction in
the spread is corroborated by the top right-hand panels of Figure
\ref{stddev.fig}, which shows the (annual) spread in distance moduli,
$\sigma$, and the errors on the means, $\varepsilon_\mu =
\sigma/\sqrt{N}$ (where $N$ is the number of measurements included in
the determination of $\sigma$) as a function of year of publication:
the $1\sigma$ spreads become significantly smaller during the period
2001--2009 compared with the pre-2001 spreads.

In addition, the two-year averages appear to cluster very tightly near
the canonical LMC distance modulus. This is exemplified in the top
left-hand panel of Figure \ref{stddev.fig}. This panel shows the
differences in the (annual) average distance moduli for the
Cepheid-based distance determinations with respect to both the
canonical value and the (annual) average values based on all distance
determinations.

\subsubsection{Cepheid-like tracers}

During this period, 37 determinations of the LMC distance modulus
based on Cepheids were published (for the full bibliography, see
http://astro-expat.info/Data/pubbias.html). Here we explore the
reason(s) for this tight clustering and the reduction in the spread of
the derived values over this period.

A detailed examination of the individual articles shows that our
database of Cepheid-based distances published during the 2001--2009
period is based---directly or indirectly---on five different
Cepheid(-like) tracers. These include (i) classical Cepheids, (ii)
bump Cepheids (three distance determinations published in three
articles by two independent teams, i.e., Bono et al. 2002a vs Keller
\& Wood 2002, 2006), (iii) triple-mode pulsators (Moskalik \&
Dziembowski 2005), (iv) Type II Cepheids (Matsunaga et al. 2009), and
(v) $\delta$ Scuti variable stars (three distance determinations by
McNamara et al. 2007). Most (i.e., 29: see our online database) of the
Cepheid-based LMC distance moduli are based on classical Cepheids, of
which one publication (Bono et al. 2002b) actually uses first-overtone
(FO) rather than fundamental-mode (FU) pulsators.

The vast majority of articles based on FU variables (22) are based on
(predominantly large) samples obtained from the Optical Gravitational
Lensing Experiment {\sc ii} ({\sc ogle ii}) photometric database
(i.e., the Cepheid sample of Udalski et al. 1999). Five determinations
in four studies (Groenewegen \& Salaris 2003; Storm et al. 2004;
Gieren et al. 2005; Testa et al. 2007) use {\sc ogle ii} observations
of the Cepheids in the LMC cluster NGC 1866 for all or part of their
analyses. Two other studies (Newman et al. 2001; Macri et al. 2006)
used the relative distance moduli between the LMC and NGC 4258,
although both teams rely on {\sc ogle ii}-based period--luminosity
relations (PLRs) to obtain the distance moduli to the individual
galaxies. This implies that a large fraction of the Cepheid-based
distance measurements in the 2001--2009 period were based on the
same---or on subsets of the same---basic photometric database, so that
these samples were not independent. The distance moduli for each of
the individual tracers are included in Table \ref{cepheids.tab}.

\begin{table}[h!]
\caption{Mean LMC distance moduli based on Cepheid-like tracers,
  2001--2009.}
\label{cepheids.tab}
\begin{center}
\begin{tabular}{@{}lcl@{}}
\hline\hline
Distance tracer & $(m-M)_0$$^a$ & Syst.  \\
                & (mag)       & error$^b$ \\
\hline
Classical Cepheids (FU)     & $18.49 \pm 0.08$ & $^c$ \\
-- NGC 4258 relative moduli & ($18.36 \pm 0.05$ & 0.26) \\
-- NGC 1866 member stars    & ($18.48 \pm 0.11$ & 0.21) \\
Classical Cepheids (FO)     & $18.51 \pm 0.03$ & 0.15 \\
Bump Cepheids               & $18.54 \pm 0.01$ & 0.10 \\
Triple-mode pulsators       & $18.44 \pm 0.10$ & 0.13$^d$ \\
Type II Cepheids            & $18.46 \pm 0.10$ & $^e$ \\
$\delta$ Scuti stars        & $18.49 \pm 0.05$ & 0.07 \\
\\
{\bf All Cepheid-like tracers} & $18.48 \pm 0.08$ & $^c$ \\
\hline\hline
\end{tabular}
\end{center}
\flushleft
$^a$ Distance moduli and their statistical uncertainties; the latter
  represent the spreads in the individual data points that contribute
  to the final values;\\
$^b$ Based on taking into account the uncertainties quoted for the
  individual published values;\\
$^c$ No systematic error determined; the distance modulus based on all
  Cepheid-like tracers is the average of all individual tracers listed
  in this table (each given equal weight), while the statistical
  uncertainty represents the spread among the individual values;\\
$^d$ Systematic uncertainty from Moskalik \& Dziembowski (2005);\\
$^e$ No systematic uncertainty available (single source).
\end{table}

\subsubsection{Calibration}

Most classical Cepheid-based distance determinations covered here, as
well as the derivations based on $\delta$ Scuti stars and Type II
Cepheids, are based on PLRs, which are characterized by a slope and a
zero-point luminosity. The zero-point calibrations in almost all
publications leading to the 2001--2009 classical Cepheid-based LMC
distance moduli are based on distances to well-understood Galactic
classical Cepheids. In turn, these were obtained on the basis of
either {\sl Hipparcos}, {\sl HST} Fine Guidance Sensor (FGS), or
interferometric parallax measurements, or Baade--Wesselink (BW)-type
calibrations. Many of the LMC calibration relations adopt PLR slopes
that are the same as those in the Milky Way. The results based on the
calibration relations used in the different articles discussed here
are thus highly correlated.

Although we cannot explicitly rule out a degree of publication bias,
we therefore conclude that the tight clustering and small spread
observed for Cepheid-based LMC distance moduli in the 2001--2009
period is most likely driven by the public availability of the {\sc
  ogle ii} Cepheid database, which was published in a convenient form
in 1999. A time lag of 1.5--2 years between this data release and the
first peer-reviewed articles based on it seems reasonable. This,
combined with easy access to the original {\sl Hipparcos} parallaxes
(ESA 1997; Perryman et al. 1997), has led to many authors using the
same zero points and {\sc ogle ii} PLR slopes. This is, hence, the
most likely main driver of the tight clustering observed for the
Cepheid-based LMC distance moduli. The {\sc ogle ii} PLR slopes are
well-defined by virtue of the large numbers of Cepheids contributing
to the relationships, which has in turn led to a reduction in the
associated uncertainties in and spread among the distance moduli.

\subsection{RR Lyrae variables}

The bottom left-hand panel of Figure \ref{lmcdist2run.fig} shows a
statistically significant clustering of distance moduli and a
pronounced reduction in the associated spread from 1999 onwards. This
follows much greater variation in both measurables during the years
prior to this period. We identified two periods during which the RR
Lyrae-based LMC distance moduli exhibit statistically significant
clustering behavior. The first period (P1) starts in 1999 and runs
through 2002, immediately followed by a second period (P2) from 2003
until the end of 2006.

\subsubsection{RR Lyrae types}

Just as for the Cepheids, the RR Lyrae-based distance tracers are also
composed of a mixture of physically different object types. Although
the majority of RR Lyrae samples considered between 1999 and 2006 are
dominated by FU RR Lyrae (RRab), FO RR Lyrae (RRc) and double-mode
pulsators (RRd) make up sizeable fractions of the most commonly used
samples: of the 18 articles with published RR Lyrae-based LMC distance
moduli in P1+P2 (see http://astro-expat.info/Data/pubbias.html),
seven\footnote{Specifically, Carretta et al. (2000), Benedict et
  al. (2002a), Clementini et al. (2003), Maio et al. (2004), Gratton
  et al. (2003), Feast (2004), and Marconi \& Clementini (2005).} are
fully or partially based on the RR Lyrae sample of Clementini et
al. (2003)---or a prepublication version of their database (e.g.,
Carretta et al. 2000; Benedict et al. 2002a)---which contains 77 RRab,
38 RRc, and 10 RRd variables. The Dall'Ora et al. (2004a,b) distance
determinations are based on a sample of 21 RRab and nine RRc
variables, while the sample of Marconi \& Clementini (2005) is
composed of 7 RRab and 7 RRc stars. In most analyses based on multiple
types of RR Lyrae pulsators, the FO pulsators were
``fundamentalized,'' i.e., an appropriate correction ($\Delta \log P
[{\rm days}] \approx +0.127$; e.g., Dall'Ora et al. 2004a,b) was
applied to their periods, so that the same
period--luminosity--metallicity ($PLZ$) relation could be used.

Of the remaining publications, Alcock et al. (2004) and Clement et
al. (2005), who share team members among their co-authors (and, hence,
most likely used similar methods; this is unclear given that the
latter reference is only an abstract), based their results on a set of
330 genuine RRc variables located near the LMC bar, which were
selected from the {\sc macho} (MAssive Compact Halo Objects)
database. Finally, Kov\'acs (2000) used 181 {\sc macho} RRd stars. We
have indicated the distance moduli resulting from the use of {\it
  only} RRc and RRd variables in the middle panels of Figure
\ref{indivdistLMC.fig}. Having assessed the RR Lyrae sample selection
criteria employed in the period between 1999 and 2006, we conclude
that there is substantial overlap among publications, which hence
renders their independence questionable.

\subsubsection{Calibration}

Second, we explored the methods used and which contribute to the
average RR Lyrae-based LMC distance moduli. These can be categorized
into three main classes, (i) those publications that use a form of the
$M_V$--[Fe/H] luminosity--metallicity relation (LMR), $\alpha [{\rm
    Fe/H}]_{\rm RR} + \beta$ (seven articles),\footnote{Specifically,
  Groenewegen \& Salaris (1999), Carretta et al. (2000), McNamara
  (2001), Clementini et al. (2003), Maio et al. (2004), Gratton et
  al. (2003), and Feast (2004).} (ii) those authors that base their
distance estimates on a $PLZ$ relation (six
articles),\footnote{Including Bono (2003), Dall'Ora et al. (2004a,b),
  Borissova et al. (2004), Rastorguev et al. (2005), and Sollima et
  al. (2006).} and (iii) those that use theoretical pulsation modeling
to derive luminosities and, hence, a distance modulus (Kov\'acs 2000;
Alcock et al. 2004; Marconi \& Clementini 2005). The zero points of
the LMRs and the $PLZ$ relations are generally based on parallax
measurements to local field or globular cluster (GC) RR Lyrae in the
Milky Way, either based on {\sl Hipparcos} or {\sl HST}
parallaxes. The general consensus among authors using an LMR is that
the slope $\alpha \approx$ [0.18--0.20] mag dex$^{-1}$ (but see
below), while most authors who relied on $PLZ$ relations adopted those
of Bono et al. (2003) and their precursors by largely the same
authors. Once again, we see that there is significant overlap between
the methods used in the publications that eventually result in the
average LMC distance moduli reported here.

Over the entire range of RR Lyrae-based distance determinations
examined, there has been a gradual shift in the preferred method from
LMRs to near-infrared (NIR) $PLZ$ relations. Empirical and theoretical
arguments indicate that LMRs might not be linear over the entire
metallicity range covered by field RR Lyrae stars (e.g., Bono et
al. 2003; Sandage \& Tammann 2006). Moreover, they are affected by
evolutionary effects such as off-zero-age HB evolution. This implies
that the mass distribution inside the RR Lyrae instability strip is
poorly known. Note that these considerations apply to both field and
cluster RR Lyrae. NIR $PLZ$ relations are minimally affected by these
problems (e.g., Del Principe et al. 2006; Sollima et al. 2006; for a
review, see Bono et al. 2011).

\subsubsection{Distance determinations based on globular cluster RR Lyrae}

Finally, we considered the spatial distributions of the RR Lyrae
samples. Twelve (P1: 4; P2: 8) of the 18 studies undertaken during the
entire period of interest explore the properties of RR Lyrae stars in
the galaxy's inner regions and/or fields in or near the LMC bar. Of
the remaining six publications, five (P1: Bono 2003; P2: Dall'Ora et
al. 2004a,b; Rastorguev et al. 2005; Sollima et al. 2006) focus on the
RR Lyrae variables in the LMC GC Reticulum. We note that the latter
studies do not necessarily provide a well-constrained distance to the
LMC's center. This is the main reason as to why we opted against
inclusion of distance moduli to {\it individual} clusters in our
compilation; Reticulum is an exception, because of its benchmark use
by a number of authors.

Only two articles venture beyond the inner galaxy or Reticulum:
Groenewegen \& Salaris (1999) and McNamara (2001) both based their LMC
distance estimate on a study of RR Lyrae stars in (the same) seven old
LMC GCs. In the middle panels of Figure \ref{indivdistLMC.fig}, we have
separately indicated the LMC distance moduli based on studies of the
Reticulum GC, as well as the GC results of Groenewegen \& Salaris
(1999) and McNamara (2001). The Reticulum-based distance
determinations do not appear to be systematically offset from the
field RR Lyrae distance estimates. This analysis strengthens us in our
conclusion that the basic samples used for RR Lyrae-based distance
determinations between 1999 and 2006 were highly
interdependent. Correlations among the resulting distance values may
thus be expected and should certainly not come as a surprise.

We also explored the possible reason(s) for the origin of the rather
sudden reduction in average distance modulus between P1 and P2. A
close examination of the middle right-hand panel of Figure
\ref{indivdistLMC.fig} reveals that most distance determinations
cluster very tightly around or just below the level of the canonical
distance modulus, with the exception of Rastorguev et al.'s (2005)
distance estimate to Reticulum. Rastorguev et al. (2005) examined 388
Galactic field RR Lyrae stars selected from the Beers et al. (2000)
catalog, as well as 1204 RRab and Rc variables with proper-motion data
located within 6 kpc from the Sun. They employed the now-outdated NIR
PLR of Jones et al. (1992), which was calibrated based on application
of the BW method, as well as statistical parallaxes to determine the
PLR's zero point. Based on the observations of Reticulum RR Lyrae
stars from Dall'Ora et al. (2004a,b), the PLR of Carney et al. (1995),
and their new zero point, they obtained a new LMC distance modulus,
which is significantly shorter than the canonical value. However, this
data point does not appreciably affect the average level during P2,
however. The scatter among the data points in the middle left-hand
panel is greater, and so are the error bars (spreads) associated with
a number of the distance moduli. The data points driving an offset of
the mean distance moduli to a level above the canonical value are
based on McNamara's (2001) distance determinations. First, we point
out that---perhaps surprisingly---the GC distance moduli of
Groenewegen \& Salaris (1999), $(m-M)_0 = 18.61 \pm 0.28$ mag, and
McNamara (2001), $(m-M)_0 = 18.61 \pm 0.04$ mag, are identical. Both
teams used LMRs, although with very different slopes: $\alpha = 0.18$
vs 0.30 mag dex$^{-1}$ (Groenewegen \& Salaris 1999 vs McNamara
2001). For the same zero point, this difference in slope leads to a
luminosity difference of 0.18 mag for the typical metallicity of LMC
RR Lyrae stars usually adopted, [Fe/H] $=-1.5$ dex, in the sense that
the LMR with the steeper slope would lead to a brighter magnitude.

The former authors used the technique of ``reduced parallax'' and
found that---for the same slope and exactly the same sample of RR
Lyrae variables---they obtained a zero point, $\beta = 0.77 \pm 0.26$
mag, which is $\sim 0.28$ mag brighter than the zero point obtained
from statistical-parallax calibration. McNamara (2001), on the other
hand, calibrated the LMR on the basis of the PLR of Galactic
high-amplitude $\delta$ Scuti stars and found $\beta = 0.92 \pm 0.09$
mag, a difference of $\Delta \beta = 0.15$ mag, which largely negates
the effect of the difference in $\alpha$ between both studies. It thus
appears that both analyses, although based on the same set of seven
LMC GCs, use independent methods to yield the same distance to the
LMC. In other words, the systematically larger LMC distance derived
from the RR Lyrae stars in these seven GCs may represent real distance
variance, or instead imply that the centroid of the GCs' distribution
is not co-located with the center of the LMC as defined by its bar.

\subsection{Features in the color--magnitude diagram}

Even a cursory glance at the bottom right-hand panel of Figure
\ref{lmcdist2run.fig} immediately shows that the LMC distance moduli
based on CMD-feature calibration converge significantly from 2006
onwards. A closer look at the publications leading to the average
values shown in this figure reveals that there is a simple reason for
this statistically significant clustering. For the period up to
2004/2005, various authors used a range of CMD features to derive
their distance moduli, including the HB level, the TRGB, the magnitude
of the helium-burning RC, as well as MS/MSTO fitting. However, with
the exception of Rubele et al.'s (2012) distance determination, since
2006 the RC magnitude has been the only diagnostic CMD feature used to
estimate the distance to the LMC.

Of the six RC-based distance determinations between 2006 and 2013 (see
our online database), most used NIR photometry of predominantly
central fields in the LMC. At NIR wavelengths, population
effects---such as those caused by differences in ages and
metallicities---are minimal, although not necessarily
negligible. [Rubele et al. (2012) also used NIR observations, which
  were newly obtained with the ESO/{\sl VISTA} telescope as part of
  the ESO public survey of the Magellanic System.] In addition, most
authors used the same set of population corrections, i.e., those
proposed by Girardi \& Salaris (2001) and Salaris \& Girardi (2002),
or fully equivalent methods. Given that the apparent magnitudes of RC
stars across the LMC can be easily deprojected to the LMC center, it
is not surprising that all recent CMD-based distance determinations
yield very similar values. It is not necessary to attribute this
effect to publication bias.

\section{Statistical analysis}
\label{statistics.sec}

\begin{table*}[b!]
\caption{``Non-compliance'' of published LMC distance moduli with our
  adopted reference values based on Freedman et al. (2001) and
  Pietrzy\'nski et al. (2013) for different tracer
  populations. Thresholds (``significance levels'') of 0.05 and 0.01
  correspond to a difference of 2 and 3 standard deviations or
  confidence intervals of 95 and 99\%, respectively.}
\label{stats2.tab}
\begin{center}
\begin{tabular}{@{}lccrrccrccrrr@{}}
\hline \hline
Difference    & Period        & \multicolumn{3}{c}{Cepheids} && \multicolumn{3}{c}{RR Lyrae} && \multicolumn{3}{c}{CMD features} \\
\cline{3-5}\cline{7-9}\cline{11-13}
w.r.t.        &               & $N_{\rm tot}$ & \multicolumn{2}{c}{Threshold} && $N_{\rm tot}$ &\multicolumn{2}{c}{Threshold} && $N_{\rm tot}$ &\multicolumn{2}{c}{Threshold} \\
\cline{4-5}\cline{8-9}\cline{12-13}
& & & 0.05 & 0.01 && & 0.05 & 0.01 && & 0.05 & 0.01 \\
\hline
Freedman      & since 01/1990 & 82 &  4 &  1 && 56 &  4 & 1 && 39 &  7 &  4 \\
et al. (2001) & since 01/2001 & 48 &  2 &  1 && 34 &  0 & 0 && 19 &  0 &  0 \\
              & since 01/2005 & 28 &  2 &  1 && 18 &  0 & 0 &&  9 &  0 &  0 \\
\hline
Pietrzy\'nski & since 01/1990 & 82 & 21 & 13 && 56 & 12 & 8 && 39 & 16 & 10 \\
et al. (2013) & since 01/2001 & 48 & 12 &  7 && 34 &  6 & 4 && 19 &  6 &  2 \\
              & since 01/2005 & 28 & 10 &  6 && 18 &  4 & 2 &&  9 &  2 &  1 \\
\hline
\hline
\end{tabular}
\end{center}
\end{table*}

To decide whether the published distance moduli show a significant
difference from the published reference values of $\mu_{\rm ref} =
18.50 \pm 0.10$ mag (Freedman et al. 2001) and $\mu_{\rm ref} = 18.493
\pm 0.008$ mag (Pietrzy\'nsky et al. 2013; statistical uncertainties
only), we perform a statistical test based on well-established
statistical principles. We regard the published distance moduli as the
mean values derived from a large population of indicators
(representing, e.g., the {\sc ogle} Cepheid or RR Lyrae samples), and
their associated standard deviations as the standard deviation of
these large populations. We assume that these values are well-defined
and disregard possible effects of sampling statistics. This approach
is justified, because all published distance moduli were derived from
studies based on large numbers of data points, so the effects of
small-number statistics can be ignored. In this framework, we can
write the test statistic as
\begin{equation}
z = \frac{\mu_{\rm dist} - \mu_{\rm ref}}{\sqrt{\sigma^2_{\rm dist} +
\sigma^2_{\rm ref}}},
\end{equation}
where $\mu_{\rm dist}$ is the published distance modulus and $\mu_{\rm
  ref}$ the reference distance modulus of interest; $\sigma_{\rm
  dist}$ corresponds to the standard deviation of the given published
distance modulus and $\sigma_{\rm ref}$ is that for the reference
distance modulus. The expression $\sigma^2_{\rm dist} + \sigma^2_{\rm
  ref}$ is the variance of the difference in means as defined by the
law of total variance for independent quantities. This expression
assumes that errors in the distance moduli follow a Gaussian
distribution. The resulting value represents a two-sided hypothesis
test, since the published values can be greater than (positive $z$
score) or less than (negative $z$ score) the reference moduli.

Using the test statistic, we compute $z$ scores for our compilation of
distance moduli and compare these scores with thresholds at the 0.05
and 0.01 levels of significance, which correspond to differences of 2
and 3 standard deviations, respectively. In other words, significances
of 0.05 and 0.01 represent confidence intervals of 95 and 99\%,
respectively. If the absolute value of a $z$ score is larger than the
threshold values at these levels of significance, then the null
hypothesis of this test, i.e., that the means are statistically equal,
is rejected, making the values statistically different at that
level. Under the frequentist-statistics interpretation, rejecting the
null hypothesis at a 0.05 or 0.01 level of significance means that the
probability that the null hypothesis is correct (hence making our
conclusion incorrect) is at most 0.05 or 0.01, respectively. The
threshold values at the 0.05 and 0.01 levels are 1.96 and 2.58,
respectively.

Table \ref{stats2.tab} provides an overview of the level of
``compliance'' of all published distance moduli for three specific
tracers (Cepheids, RR Lyrae, and CMD features) with the reference
distance moduli over wide intervals of publication date, coinciding
with the intervals adopted for the analysis presented in Table
\ref{stats.tab}. First, we note that the ``non-compliance rate'' with
respect to the Pietrzy\'nski et al. (2013) reference value is
significantly higher than that with respect to the ``canonical''
distance modulus of Freedman et al. (2001). This can be ascribed
entirely as owing to the much smaller statistical uncertainty
associated with the former value. Second, we point out that since
2001, the published distance moduli based on all three tracer
populations have been similar to Freedman et al.'s (2001) value at
better than the 0.05 level of significance; for normally distributed
data points, this corresponds to the $2\sigma$ level or 95\%
confidence. Although the Pietrzy\'nski et al. (2013) result has much
smaller statistical uncertainties, the general trend seen is the same:
the LMC distance moduli published since 2001 are very well represented
by the recommended value of $(m-M)_0 = 18.49$ mag (cf. Section
\ref{conclusions.sec}).

It is interesting to explore which of the published distance moduli
are significantly different, at (much) worse than the 95\% confidence
level (i.e., worse than the 0.05 level of significance). Only a small
number of published distance moduli deviate significantly from the
reference value of Freedman et al. (2001), which is associated with
the larger (statistical) uncertainty. Haschke et al.'s (2012) LMC
distance modulus of $(m-M)_0 = 18.85 \pm 0.08$ mag based on Cepheid
observations and adopting an area-averaged reddening value is given a
$z$ score of 2.73. However, once they correct for the reddening using
values specific to each individual Cepheid variable in their sample,
the resulting distance modulus of $(m-M)_0 = 18.65 \pm 0.07$ mag is no
longer flagged up as an outlier by our test statistic. The LMC
distance modulus of Walker et al. (2001), $(m-M)_0 = 18.33 \pm 0.05$
mag, attracts a $z$ score of $-3.22$. This distance modulus was
obtained based on CMD fits to NGC 1866, under the assumption that the
cluster would be located in the LMC disk plane. (Releasing that latter
assumption would increase the cluster's distance modulus to 18.35
mag.) We note that Salaris et al. (2003), Groenewegen \& Salaris
(2003), and Storm et al. (2006) derive very similar distance moduli to
the cluster.

Finally, application of our test statistic to the distance moduli
obtained from analysis of the geometric distance indicators, EB
systems, and SN 1987A confirms our earlier conclusions. For the EB
distances, the values of Guinan et al. (1998) and Fitzpatrick et
al. (2003) disagree with the Pietrzy\'nski et al. (2013) reference
value at worse than the 0.01 level of significance; the Udalski et
al. (1998) result differs at worse than the 0.05 level of
significance. There are no EB differences with respect to the Freedman
et al. (2001) value at the 0.01 level of significance, but the
Fitzpatrick et al. (2003) result differs at the 0.05 level of
significance. We already discussed the main physical reasons for these
differences in Section \ref{conclusions.sec}; the same reasons apply
to the Udalski et al. (1998) determination.

As regards the SN 1987A-based distance moduli, they all agree with the
Freedman et al. (2001) value at the 99\% confidence level, although a
number of distance determinations thus derived disagree with the more
recent reference value at that level: of the 15 distance moduli
published based on analyses of SN 1987A, Gould (1995), Eastman et
al. (1996), Lundqvist \& Sonneborn (1997), and Gould \& Uza (1998)
published distance moduli that yield $z$ scores of, respectively,
$-3.02, -5.38, 3.50$, and $-3.02$ with respect to the Pietrzy\'nski et
al. (2013) reference value. Note that the high negative $z$ score of
the Eastman et al. (1996) result is artificially inflated because
these authors did not report an uncertainty on their estimate. The
other values do not show a preference for a systematic difference, so
we are left to conclude that the high $z$ scores may be related to the
underlying assumptions made.

\section{Verdict}
\label{conclusions.sec}

\begin{figure}
\begin{center}
\includegraphics[width=\columnwidth]{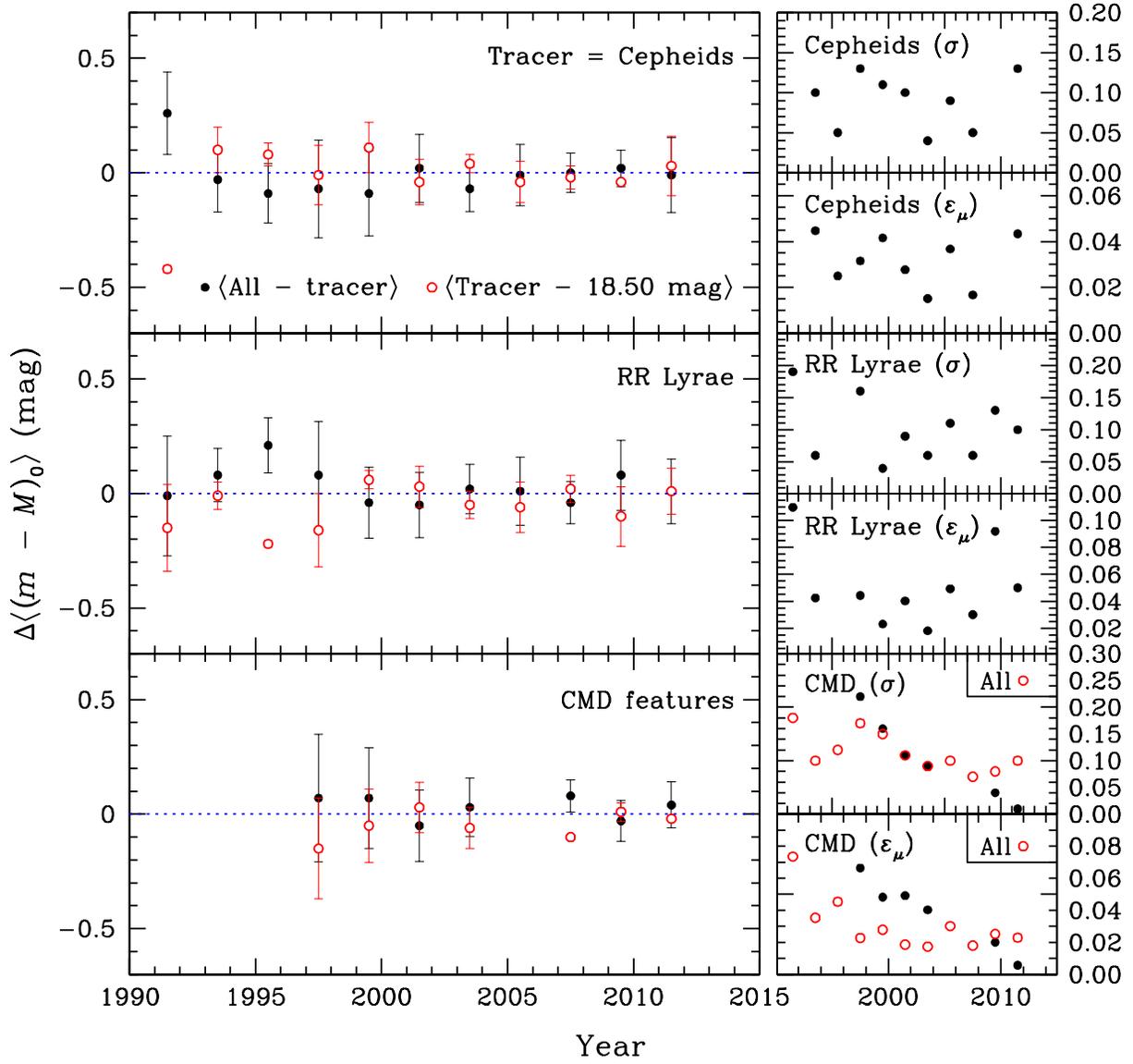}
\caption{(left) Differences in distance moduli for three common
  distance tracers. Black bullets: mean differences between all
  published distance moduli for a given period and those for a
  specific tracer. Red open circles: mean differences between the
  distance moduli for a given tracer and the canonical distance
  modulus. (right) Mean spreads in the data values, $\sigma$, and
  errors on the means, $\varepsilon_\mu = \sigma/\sqrt{N}$, as a
  function of date of publication for the three common distance
  tracers and (bottom right-hand panels, red open circles) for all
  measurements. The units of $\sigma$ and $\varepsilon_\mu$ in the
  right-hand panels are given in magnitudes.}
\label{stddev.fig}
\end{center}
\end{figure}

In the previous sections, we made a case of highly correlated results
among populations for a given distance tracer, rather than attributing
the observed tight clustering of the distance moduli and the clear
reduction in the associated spreads to publication bias.\footnote{We
  specifically point out that most of these conclusions pertain to
  very similar time intervals as that considered by Schaefer (2008),
  yet we do not concur with his conclusions.} Here, we will
additionally explore the behavior of the {\it spread} in the distance
moduli and the errors on the means as a function of publication
date. Figure \ref{stddev.fig} (left-hand panels) shows the
differences, averaged over two-year intervals, of the Cepheid-, RR
Lyrae-, and CMD-based distance moduli with respect to both the
canonical distance modulus of 18.50 mag and the ensemble of all
determinations published during the relevant time spans. In almost all
cases, the two-year average distance moduli are consistent with the
canonical distance modulus within the $1\sigma$ spreads (as indicated
by the red open circles), in particular since the late 1990s, i.e.,
well before Freedman et al.'s (2001) seminal publication. Again,
combined with our analysis in Section \ref{bias.sec}, this leads us to
suggest that the observed trend(s) may be largely driven by the common
availability of the {\sc ogle ii} data and the community's easy access
to the full {\sl Hipparcos} parallax database.

The right-hand panels of Figure \ref{stddev.fig} show the spreads and
the errors on the means among the distance moduli as a function of
publication date. For the Cepheid- and RR Lyrae-based distance moduli,
there are no clearly discernable trends. However, the uncertainties
are increasingly reduced as a function of increasing publication date
for the CMD-based distances over the entire period covered. The bottom
right-hand panels additionally include the same trend for the full
ensemble of distance measurements (labeled ``All''). The full sample
of distance moduli also shows an obvious decrease in the spread
between 1998 and 2008. This trend is clearly driven by the reduction
in the uncertainties associated with the CMD-based measurements. In
turn, this is most likely due to the improved local calibration of the
absolute RC magnitude (and as a function of wavelength).

So far, we have focused on the results based on a number of
well-represented individual distance tracers. However, combined
analyses can be used to reinforce the results obtained from different
tracers. We concur with Schaefer (2008) that this might indeed lead to
a degree of publication bias, because it is intrinsic to human nature
to publish results that conform to some extent to the norm, while
strongly deviating results may not see the light of day in
peer-reviewed publications. On the other hand, cross checking the
results obtained from a variety of tracers can also help to prevent
such effects from occurring in the first place, in particular if the
techniques employed are independent (for a discussion, including a
historical perspective, see Tammann et al. 2008). For instance, Laney
et al. (2012) used the LMC's $H$- and $K$-band RC magnitudes to derive
LMC distance moduli, with and without population corrections, and
concluded that the very small differences found as a function of NIR
wavelength ``imply that any correction to the $K$-band Cepheid PL[R]
due to metallicity differences between Cepheids in the LMC and the
solar neighborhood must be quite small.''

Where does this leave us in the context of using the LMC's distance as
a key rung of the local extragalactic distance ladder? While we
conclude that the effects of publication bias may have been overplayed
in previous publications, the resulting LMC distance moduli are likely
still affected by poorly understood systematic uncertainties. This is
exemplified by the increasing spreads and the variation in the mean
distance modulus in the more recent past, in particular as pertaining
to the Cepheid variables: see Figure \ref{lmcdist2run.fig} (top
right-hand panel). Similarly, the sustained variation in the mean
two-year running distance moduli based on RR Lyrae variables (see
Figure \ref{lmcdist2run.fig}, bottom left-hand panel), combined with
the sizeable standard deviations and our much improved understanding
of the physical processes dominating RR Lyrae variability, supports
the persistence of systematic uncertainties.

Fortunately, the LMC is located sufficiently nearby that a few types
of geometric distance tracers are fairly readily available, including
the enigmatic SN 1987A's ring (a light echo) and an ever increasing
number of EB systems. While results based on studies using the SN
1987A ring should be treated with caution (cf. Gould \& Uza 1998; de
Grijs 2011, his chapter 3.7.2), significant progress has recently been
made in the use of EBs in the LMC as high-precision distance
anchors. Pietrzy\'nski et al. (2013) determined the direct distances
to eight long-period, late-type EB systems in the LMC. Their resulting
distance modulus, $(m-M)_0= 18.493 \pm 0.008 \pm 0.047$ mag (where the
first and second uncertainties refer to the statistical and systematic
errors, respectively), is accurate to 2.2\% and confirms that the
canonical distance modulus of $(m-M)_0= 18.50 \pm 0.10$ is indeed a
very reasonable approximation.

Pietrzy\'nski et al. (2013) make a convincing case for the use of
long-period cool giant stars, because one can accurately measure both
the linear and the angular sizes of their components. This presents
more problems for hot, early-type systems. However, Schaefer (2013)
calls their result into question; in addition to repeating his
concerns regarding a bandwagon effect, he points out that
Pietrzy\'nski et al.'s (2013) LMC distance of $D_{\rm LMC} = 49.97 \pm
0.19 \mbox{ (stat.)} \pm 1.11$ (syst.) kpc is significantly different
from the average distance of four hot, early-type EBs, $D_{\rm LMC} =
47.1 \pm 1.4$ kpc, published by Guinan et al. (1998), Fitzpatrick et
al. (2002, 2003), and Ribas et al. (2002). This systematic difference
between the results of both groups is most likely an indication of
systematic uncertainties, which would particularly affect the earlier
results. These were based on systems composed of hot early-type stars,
which implies that these authors were limited to using theoretical
models (energy distributions); accounting for the systematic
uncertainties associated with this approach is notoriously
difficult. Pietrzy\'nski et al. (2013) discovered giant stars in EB
systems, which uniquely allowed them to use the very well calibrated
surface-brightness--$(V-K)$ color relation for such stars to determine
their angular sizes, so that their error estimates are more robust and
less affected by lingering systematic effects (G. Pietrzy\'nski,
priv. comm.).

In relation to this issue, Schaefer's (2013) comment actually ignores
the recent EB results from one independent study (Bonanos et al. 2011)
and one by Prada Moroni et al. (2012), composed of a subset of authors
contributing to the Pietrzy\'nski et al. (2013) results (see Figure
\ref{lmcdist.fig}, bottom right-hand panel). Both implied a
systematically greater distance to the LMC than the 1998--2003
counterexamples produced by Schaefer (2013). In addition, one of the
most recently published LMC distance moduli in our database (Marconi
et al. 2013) is based on theoretical pulsation modeling of the light
and radial velocity curves of an EB; it also yields a distance that is
in good agreement with that derived by Pietrzy\'nski et al. (2013).

Schaefer (2013) also points out that Pietrzy\'nski's team had
previously reported the distance to one of its EB systems,
OGLE-LMC-ECL-09114, as $D= 50.1 \pm 1.4$ kpc (Pietrzy\'nski et
al. 2009), but their updated analysis leads to a new estimate of $D =
49.3 \pm 0.5$ kpc. We note that the earlier value was associated with
a large error bar, $(m-M)_0 = 18.50 \pm 0.55$ mag. The small change in
system parameters determined by Pietrzy\'nski et al. (2013), which is
is well within those $1\sigma$ uncertainties, is driven by a
significant body of new observations obtained during the system's
eclipses (G. Pietrzy\'nski, priv. comm.). This allowed these authors
to improve their model and also take into account limb darkening and
other secondary effects, as shown in their supplementary table 5.

\begin{table}[h!]
\caption{Statistical properties of the body of LMC distance
  measurements for two representative recent periods. Means and
  population standard deviations are given in units of magnitudes.}
\label{stats.tab}
\begin{center}
\begin{tabular}{@{}llcc@{}}
\hline \hline
             &          & \multicolumn{2}{c}{Period} \\
\cline{3-4}
             &          & 01/2001-- & 01/2005-- \\
             &          & 12/2013   & 12/2013   \\
\hline
             & Mean     & 18.481 & 18.486 \\
All          & $\sigma$ &  0.097 &  0.091 \\
             & $N$      & 125    & 68     \\
\hline
             & Mean     & 18.486 & 18.484 \\
Cepheids     & $\sigma$ &  0.100 &  0.107 \\
             & $N$      & 48     & 28     \\
\hline
             & Mean     & 18.472 & 18.471 \\
RR Lyrae     & $\sigma$ &  0.094 &  0.103 \\
             & $N$      & 38     & 22     \\
\hline
             & Mean     & 18.499 & 18.505 \\
CMD features & $\sigma$ &  0.153 &  0.133 \\
             & $N$      & 19     &  9     \\
\hline
{\bf Weighted} & {\bf Mean} & {\bf 18.492} & {\bf 18.492} \\
             & {\bf $\sigma$} & {\bf 0.089} & {\bf 0.090} \\
\hline \hline
\end{tabular}
\end{center}
\end{table}

In view of our analysis and discussion of the statistically
significant clustering of LMC distance moduli and the observed
reduction in their spread over the past two decades, we conclude that
strong publication bias is unlikely to have affected the majority of
published LMC distance moduli. Note that our conclusions pertain to
the largest, most complete data set of LMC distance moduli available
to date, superseding all previous compilations. To get a handle on the
systematic spread in distance determinations owing to, e.g., depth
effects and real cosmic variance, we determined the average distance
moduli and the associated population standard deviations for all
published distance determinations, as well as for those based on
Cepheid and RR Lyrae variables and on CMD features, for two recent
periods, i.e. since 2001 and since 2005 until the present time. The
results are shown in Table \ref{stats.tab}. The final section of Table
\ref{stats.tab} lists the weighted means and standard deviations based
on a careful combination, in a statistical sense (i.e., following
Watkins et al. 2013, their Appendix A), of the mean distance moduli
and their uncertainties for the Cepheid and RR Lyrae variables, the
CMD features, and the latest EBs result of Pietrzy\'nski et
al. (2013). Our main underlying assumption adopted here was that the
individual data points are distributed normally (i.e., approximately
in a Gaussian fashion); this is a good first-order approximation to
all data sets pertaining to the three specific tracers included in the
table.

For a more in-depth statistical analysis, we refer the reader to
Section \ref{statistics.sec}. In view of these results, as well as
those listed in Table \ref{cepheids.tab}, we recommend that a slightly
updated canonical distance modulus of $(m-M)_0 = 18.49 \pm 0.09$ mag
be used for all practical purposes that require a general distance
scale without the need for accuracies better than a few percent. In
Paper II in this series, we extend our analysis to the body of
distance measurements for M31, M33, and a number of their companion
galaxies, and place our recommendations in the context of distance
measures to Local Group galaxies and beyond.

\begin{acknowledgements}
RdG is grateful for research support from the National Natural Science
Foundation of China through grants 11073001 and 11373010. This work
was partially supported by PRIN--INAF 2011 (PI M. Marconi) and by
PRIN--MIUR (2010LY5N2T, PI F. Matteucci). GB thanks the Carnegie
Observatories for support as a science visitor. We thank Grzegorz
Pietrzy\'nski for clarifying a number of issues related to their 2013
geometric distance determination to the LMC. We also acknowledge
contributions by David Schlachtberger during the early stages of this
project. This research has made extensive use of NASA's Astrophysics
Data System Abstract Service.
\end{acknowledgements}


\begin{thebibliography}{}

\bibitem[]{} Alcock, C., et al. 2004, AJ, 127, 334
\bibitem[]{} Bailer-Jones, C. A. L. 2009, Int'l J. Astrobiol., 8, 213
\bibitem[\protect\citeauthoryear{Beers et
    al.}{2000}]{2000AJ....119.2866B} Beers, T.~C., Chiba, M., Yoshii,
  Y., Platais, I., Hanson, R.~B., Fuchs, B., \& Rossi S. 2000, AJ,
  119, 2866
\bibitem[]{} Begg, C. B., \& Berlin, J. A. 1988, J. R. Stat. Soc. A,
  151, 419
\bibitem[]{} Bellazzini, M., Ferraro, F. R., Sollima, A., Pancino, E.,
  \& Origlia, L. 2004, A\&A, 424, 199
\bibitem[\protect\citeauthoryear{Benedict et
    al.}{2002a}]{2002AJ....123..473B} Benedict, G.~F., et al. 2002a,
  AJ, 123, 473
\bibitem[]{} Benedict, G. F., et al. 2002b, AJ, 124, 1695
\bibitem[]{} Bonanos, A. Z., Castro, N., Macri, L. M., \& Kudritzki,
  R.-P.  2011, ApJ, 729, L9
\bibitem[]{} Bono, G. 2003, Lect. Notes Phys., 635, 85
\bibitem[\protect\citeauthoryear{Bono et
    al.}{2001}]{2001MNRAS.326.1183B} Bono, G., Caputo, F., Castellani,
  V., Marconi, M., \& Storm, J. 2001, MNRAS, 326, 1183
\bibitem[]{} Bono, G., Castellani, V., \& Marconi M. 2002a, ApJ, 565,
  L83
\bibitem[Bono et al.(2011)]{2011rrls.conf....1B} Bono, G., Dall'Ora,
  M., Caputo, F., et al.\ 2011, in: RR Lyrae Stars, Metal-Poor Stars,
  and the Galaxy, ed. A. McWilliam, Carnegie Obs. Astrophys. Ser., 5,
  1
\bibitem[]{} Bono, G., Groenewegen, M. A. T., Marconi, M., Caputo,
  F. 2002b, ApJ, 574, L33
\bibitem[\protect\citeauthoryear{Bono et
    al.}{2003}]{2003MNRAS.344.1097B} Bono, G., Caputo, F., Castellani,
  V., Marconi, M., Storm, J., \& Degl'Innocenti, S. 2003, MNRAS, 344,
  1097
\bibitem[]{} Borissova, J., Minniti, D., Rejkuba, M., Alves, D., Cook,
  K. H., \& Freeman, K. C. 2004, A\&A, 423, 97
\bibitem[]{} Borissova, J., Rejkuba, M., Minniti, D., Catelan, M., \&
  Ivanov, V. D. 2009, A\&A, 502, 505
\bibitem[]{} Caputo, F. 1997, MNRAS, 284, 994
\bibitem[\protect\citeauthoryear{Carney et
    al.}{1995}]{1995AJ....110.1674C} Carney, B.~W., Fulbright, J.~P.,
  Terndrup, D.~M., Suntzeff, N.~B., \& Walker, A.~R. 1995, AJ, 110,
  1674
\bibitem[]{} Carretta, E., Gratton, R. G., Clementini, G., \& Fusi
  Pecci, F.  2000, ApJ, 533, 215
\bibitem[]{} Clement, C. M., Xu, X., \& Muzzin, A. V. 2005,
  Bull. Am. Astron. Soc., 37, 1364
\bibitem[\protect\citeauthoryear{Clementini et
    al.}{2003}]{2003AJ....125.1309C} Clementini, G., Gratton, R.,
  Bragaglia, A., Carretta, E., Di Fabrizio, L., \& Maio, M. 2003, AJ,
  125, 1309
\bibitem[]{} Dall'Ora, M., et al. 2004a, Mem. Soc. Astron. It., 75,
  138
\bibitem[]{} Dall'Ora, M., et al. 2004b, ApJ, 610, 269
\bibitem[]{} Dambis, A.~K., Berdnikov, L.~N., Kniazev, A.~Y., et
  al.\ 2013, MNRAS, 435, 3206
\bibitem[]{} de Grijs, R. 2011, {\it An Introduction to Distance
  Measurement in Astronomy}, Wiley-Blackwell Acad. Publ. 
\bibitem[]{} de Grijs, R., Wicker, J. E., \& Bono, G. 2014, AJ,
  submitted (Paper II)
\bibitem[Del Principe et al.(2006)]{2006ApJ...652..362D} Del Principe,
  M., Piersimoni, A.~M., Storm, J., et al.\ 2006, ApJ, 652, 362
\bibitem[]{} Di Benedetto, G. P. 1994, A\&A, 285, 819
\bibitem[Eastman et al.(1996)]{1996ApJ...466..911E} Eastman, R.~G.,
  Schmidt, B.~P., \& Kirshner, R. 1996, ApJ, 466, 911
\bibitem[]{} ESA 1997, The Hipparcos and Tycho Catalogues, ESA SP-1200
\bibitem[]{} Feast, M. 1995, in: ASP Conf. Ser., Astrophysical
  applications of stellar pulsation, eds. R. S. Stobie \&
  P. A. Whitelock, vol. 83, p. 209
\bibitem[]{} Feast, M. 1999, PASP, 111, 775
\bibitem[]{} Feast, M. 2004, in: Science with SALT, unpublished
  (astro-ph/0405015)
\bibitem[]{} Feigelson, E., \& Babu, G. J. 2013, {\it Beware the
  Kolmogorov-Smirnov test!},
  https://asaip.psu.edu/Articles/beware-the-kolmogorov-smirnov-test
  (accessed 5 July 2013)
\bibitem[\protect\citeauthoryear{Fitzpatrick et
    al.}{2002}]{2002ApJ...564..260F} Fitzpatrick, E.~L., Ribas, I.,
  Guinan, E.~F., DeWarf, L.~E., Maloney, F.~P., \& Massa, D. 2002,
  ApJ, 564, 260
\bibitem[\protect\citeauthoryear{Fitzpatrick et
    al.}{2003}]{2003ApJ...587..685F} Fitzpatrick, E.~L., Ribas, I.,
  Guinan, E.~F., Maloney, F.~P., \& Claret, A. 2003, ApJ, 587, 685
\bibitem[\protect\citeauthoryear{Foley et
    al.}{2012}]{2012ApJ...752..101F} Foley, R.~J., et al. 2012, ApJ,
  752, 101
\bibitem[]{} Freedman, W. L., et al. 2001, ApJ, 553, 47
\bibitem[\protect\citeauthoryear{Gibson}{2000}]{2000MmSAI..71..693G}
  Gibson, B.~K. 2000, Mem. Soc. Astron. It., 71, 693
\bibitem[]{} Gieren, W., Storm, J., Barnes {\sc iii}, T. G., Fouqu\'e,
  P., Pietrzy\'nski, G., \& Kienzle, F. 2005, ApJ, 627, 224
\bibitem[]{} Girardi, L., \& Salaris, M. 2001, MNRAS, 323, 109
\bibitem[Gould(1995)]{1995ApJ...452..189G} Gould, A. 1995, ApJ, 452,
  189
\bibitem[]{} Gould, A., \& Uza, O. 1998, ApJ, 494, 118
\bibitem[]{} Gratton, R. G. 1998, MNRAS, 296, 739
\bibitem[]{} Gratton, R. G., Bragaglia, A., Carretta, E., Clementini,
  G., Desidera, S., Grundahl, F., \& Lucatello, S. 2003, A\&A, 408,
  529
\bibitem[]{} Groenewegen, M. A. T., \& Oudmaijer, R. D. 2000, A\&A,
  356, 849
\bibitem[]{} Groenewegen, M. A. T., \& Salaris, M. 1999, A\&A, 348, L33
\bibitem[]{} Groenewegen, M. A. T., \& Salaris, M. 2003, A\&A, 410, 887
\bibitem[\protect\citeauthoryear{Guinan et
    al.}{1998}]{1998ApJ...509L..21G} Guinan, E.~F., et al. 1998, ApJ,
  509, L21
\bibitem[]{} Haschke, R., Grebel, E. K., \& Duffau, S. 2011, AJ, 141,
  158
\bibitem[]{} Haschke, R., Grebel, E. K., \& Duffau, S. 2012, AJ, 144,
  106
\bibitem[]{} Inno, L., et al. 2013, ApJ, 764, 84
\bibitem[\protect\citeauthoryear{Jones et
    al.}{1992}]{1992ApJ...386..646J} Jones, R.~V., Carney, B.~W.,
  Storm, J., \& Latham, D.~W. 1992, ApJ, 386, 646
\bibitem[]{} Keller, S. C., \& Wood, P. R. 2002, ApJ, 578, 144
\bibitem[]{} Keller, S. C., \& Wood, P. R. 2006, ApJ, 642, 834
\bibitem[]{} Kerber, L. O., Santiago, B. X., Castro, R., \&
  Valls-Gabaud, D.  2002, A\&A, 390, 121
\bibitem[]{} Kov\'acs, G. 2000, A\&A, 363, L1
\bibitem[]{} Laney, C. D., Joner, M. D., \& Pietrzy\'nski, G. 2012,
  MNRAS, 419, 1637
\bibitem[\protect\citeauthoryear{Liddle}{2004}]{2004MNRAS.351L..49L}
  Liddle, A.~R. 2004, MNRAS, 351, L49
\bibitem[]{} Lundqvist, P., \& Sonneborn, G. 1997, unpublished
  (astro-ph/9707144)
\bibitem[]{} Luri, X., Gomez, A. E., Torra, J., Figueras, F., \&
  Mennessier, M. O. 1998, A\&A, 335, L81
\bibitem[]{} Macri, L. M., Stanek, K. Z., Bersier, D., Greenhill,
  L. J., \& Reid, M. J. 2006, ApJ, 652, 1133
\bibitem[]{} Madore, B. F., \& Freedman, W. L. 1998, ApJ, 492, 110
\bibitem[]{} Maio, M., et al. 2004, Mem. Soc. Astron. It., 75, 130
\bibitem[]{} Marconi, M., \& Clementini, G. 2005, AJ, 129, 2257
\bibitem[]{} Marconi, M., et al. 2013, ApJ, 768, L6
\bibitem[]{} Matsunaga, N., Feast, M. W., \& Menzies, J. W. 2009, AIP
  Conf. Proc., 1170, 96
\bibitem[]{} McConnachie, A. W., Irwin, M. J., Ferguson, A. M. N.,
  Ibata, R. A., Lewis, G. F., \& Tanvir, N. 2005, MNRAS, 356, 979
\bibitem[]{} McNamara, D. H. 2001, PASP, 113, 335
\bibitem[]{} McNamara, D. H., Clementini, G., \& Marconi, M. 2007, AJ,
  133, 2752
\bibitem[]{} Moskalik, P., \& Dziembowski, W. A. 2005, A\&A, 434, 1077
\bibitem[]{} Naylor, C. 1997, Br. Med. J., 315, 617
\bibitem[]{} Newman, J. A., Ferrarese, L., Stetson, P. B., Maoz, E.,
  Zepf, S. E., Davis, M., Freedman, W. L., \& Madore, B. F. 2001, ApJ,
  553, 562
\bibitem[]{} Ngeow, C., \& Kanbur, S. M. 2008, in: Galaxies in the
  Local Volume, Astrophys. Space Sci. Proc., p. 317
\bibitem[]{} Perryman, M. A. C., et al. 1997, A\&A, 323, L49
\bibitem[\protect\citeauthoryear{Pietrzy{\'n}ski et
    al.}{2009}]{2009ApJ...697..862P} Pietrzy{\'n}ski, G., et al. 2009,
  ApJ, 697, 862
\bibitem[]{} Pietrzy\'nski, G., et al. 2013, Nat, 495, 76
\bibitem[]{} Popowski, P., \& Gould, A. 1998, ApJ, 506, 259
\bibitem[]{} Popowski, P., \& Gould, A. 1999, in: Post-Hipparcos Cosmic
  Candles, eds. A. Heck \& F. Caputo (Dordrecht: Kluwer), p. 53
\bibitem[]{} Prada Moroni, P. G., Gennaro, M., Bono, G., Pietrzy\'ski,
  G., Gieren, W., Pilecki, B., Graczyk, D., \& Thompson, I. B. 2012,
  ApJ, 749, 108
\bibitem[]{} Rastorguev, A. S., Dambis, A. K., \& Zabolotskikh, M. V.
  2005, in: The Three-Dimensional Universe with Gaia (ESA SP-576),
  eds. C. Turon, K. S. O'Flaherty, \& M. A. C. Perryman, p. 707
\bibitem[]{} Reid, N. 1998, AJ, 115, 204
\bibitem[\protect\citeauthoryear{Ribas et
    al.}{2002}]{2002ApJ...574..771R} Ribas, I., Fitzpatrick, E.~L.,
  Maloney, F.~P., Guinan, E.~F., \& Udalski, A. 2002, ApJ, 574, 771
\bibitem[]{} Ripepi, V., et al. 2013, MNRAS, 424, 1807
\bibitem[]{} Romaniello, M., Salaris, M., Cassisi, S., \& Panagia,
  N. 2000, ApJ, 530, 738
\bibitem[]{} Rosenthal, R. 1979, Psychol. Bull., 86, 638
\bibitem[]{} Rubele, S., et al. 2012, A\&A, 537, A106
\bibitem[]{} Sakai, S., Zaritsky, D., \& Kennicutt Jr., R. C. 2000,
  AJ, 119, 1197
\bibitem[]{} Salaris, M., \& Cassisi, S. 1997, MNRAS, 289, 406
\bibitem[]{} Salaris, M., \& Girardi, L. 2002, MNRAS, 337, 332
\bibitem[]{} Salaris, M., Percival, S., Brocato, E., Raimondo, G., \&
  Walker, A. R. 2003, ApJ, 588, 801
\bibitem[]{} Sandage, A., \& Tammann, G. 2006, ARA\&A, 44, 93
\bibitem[]{} Schaefer, B. E. 2008, AJ, 135, 112
\bibitem[]{} Schaefer, B. E. 2013, Nat, 495, 51
\bibitem[]{} Slosar, A., \& Seljak, U. 2004, Phys. Rev. D, 70, 083002
\bibitem[]{} Slosar, A., Seljak, U., \& Makarov, A. 2004,
  Phys. Rev. D, 69, 123003
\bibitem[]{} Sollima, A., Cacciari, C., \& Valenti, E. 2006, MNRAS,
  372, 1675
\bibitem[]{} Steer, I. P., \& Madore, B. F. 2007, {\it A Compilation of
  over 200 Published Distances to the Large Magellanic Cloud},
  http://ned.ipac.caltech.edu/level5/NED0D/LMC$_{-}$ref.html (version
  4 May 2007; accessed 5 July 2013)
\bibitem[]{} Sterling, T. D. 1959, J. Am. Stat. Assoc., 54, 30 
\bibitem[]{} Stern, J. M., \& Simes, R. J. 1997, Br. Med. J., 315, 640
\bibitem[]{} Sternberg, A., et al. 2011, Science, 333, 856
\bibitem[]{} Sterne, J. A. C., Gavaghan, D., \& Egger, M. 2000,
  J. Clin.  Epidemiol., 53, 1119
\bibitem[]{} Storm, J., Carney, B. W., Gieren, W. P., Fouqu\'e, P.,
  Latham, D. W., \& Fry, A. M. 2004, A\&A, 415, 531
\bibitem[]{} Storm, J., Gieren, W., Fouqu\'e, P., Barnes {\sc iii},
  T. G., \& G\'omez, M. 2006, Mem. Soc. Astron. It., 77, 261
\bibitem[]{} Subramanian, S., \& Subramaniam, A. 2010, A\&A, 520, A24
\bibitem[]{} Tammann, G. A., Sandage, A., \& Reindl, B. 2008, ApJ,
  679, 52
\bibitem[]{} Testa, V., et al. 2007, A\&A, 462, 599
\bibitem[]{} Udalski, A. 1998, Acta Astron., 48, 113
\bibitem[Udalski et al.(1998)]{1998ApJ...509L..25U} Udalski, A.,
  Pietrzy{\'n}ski, G., Wo{\'z}niak, P., Szyma\'nski, M., Kubiak, M.,
  \& $\dot{\rm Z}$ebru\'n, K. 1998, ApJ, 509, L25
\bibitem[\protect\citeauthoryear{Udalski et
    al.}{1999}]{1999AcA....49..223U} Udalski, A., Soszy\'nski, I.,
  Szyma\'nski, M., Kubiak, M., Pietrzy\'nski, G., Wo\'zniak, P., \&
  $\dot{\rm Z}$ebru\'n, K. 1999, Acta Astron., 49, 223
\bibitem[]{} van Leeuwen, F., Feast, M. W., Whitelock, P. A., \&
  Yudin, B.  1997, MNRAS, 287, 955
\bibitem[]{} Vaughan, S., \& Uttley, P. 2008, MNRAS, 390, 421
\bibitem[]{} Walker, A. 1999, in: Post-Hipparcos Cosmic Candles,
  eds. A. Heck \& F. Caputo (Dordrecht: Kluwer), p. 125
\bibitem[]{} Walker, A. R. 2012, Ap\&SS, 341, 43
\bibitem[]{} Walker, A. R., Raimondo, G., Di Carlo, E., Brocato, E.,
  Castellani, V., \& Hill, V. 2001, ApJ, 560, L139
\bibitem[]{} Watkins, L. L., Evans, N. W., \& van de Ven, G. 2013,
  MNRAS, 430, 971

\end{thebibliography}
\end{document}